# DYNAMIC LOCALIZATION AND ELECTROMAGNETIC TRANSPARENCY OF SEMICONDUCTOR SUPERLATTICE IN MULTIFREQUENCY ELECTRIC FIELDS

J.Y. ROMANOVA[*], Y.A. ROMANOV

Institute for Physics of Microstructures RAS, Nizhny Novgorod, 603950, Russia

**Abstract**

We consider the dynamics of electrons in semiconductor superlattices in intense multifrequency electric fields. We examine the conditions for dynamic localization and electromagnetic transparency. We investigate processes of formation, destruction and stabilization of electromagnetic transparency in biharmonic field. The stable dc fields are founded.

## 1. Introduction

In the semiconductor superlattice (SL) with relatively narrow Brillouin minizone electrons perform complex Bloch oscillations (BO) under strong electric fields. For specific ratio of amplitude and frequency of applied harmonic field the value of average electron velocity (no scattering) vanishes independently of its initial momentum. This effect is called dynamic localization (DL) in literature. One of macroscopic manifestation of complex Bloch oscillations is nonmonotonic dependence of high-frequency conductivity on amplitudes and frequencies of fields. In particular, the effects of induced (IT) and self-induced transparencies (SIT) arise.

The conditions for SIT within the single $\tau$- approximation for a one-dimensional superlattice sample with harmonic dispersion law are the same as for dynamic localization, but these effects have different physical origins [1, 2]. As a rule transparency state is unstable to the generation of both static and additional harmonic fields. There are some scenarios of destruction of SIT. Which of them will dominate depended on which process - the dc field or the high-frequency field generation and amplification will proceeds with a faster-growing increment. In reality situation internal SL field is multifrequency field even if the external field is harmonic. So for the better understanding of electromagnetic properties of the SL we have to investigate the behavior of SL in given multifrequency electric fields.

In the present paper we examine the effects of DL, electromagnetic transparency and current and voltage self - oscillations in SL, which arise under multifrequency fields. We used Boltzmann kinetic equation within $\tau$- approximation ($\tau$ is the relaxation time of the electron momentum distribution) and equation, which describe continuity of the total current in the broken circuit (the dc current is zero) [3, 4]. Our model applies to SL with arbitrary minizone dispersion law.

## 2. Dynamic Localization

We represent the longitudinal energy $\varepsilon_3$ $(k_3)$ in the form of the Fourier series (in the form of a sum of partial sinusoidal minibands) as

$$\varepsilon_3(k_3) = \sum_{n=1}^{N} \varepsilon_n(k_3) = \frac{1}{2}\sum_{n=1}^{N} \Delta_n [1 - \cos(nk_3 d)] , \qquad (1)$$

where $\Delta_n$ is width of $n$-th partial sinusoidal miniband, $d$ is a period of SL. This width may be both positive and negative. It defined by overlapping of wave functions of electron states (Wannier functions), which are localized on ν-th and *(ν+k)-th* wells (see Ref.5 for example). *N* is integer and defines the maximum number of nearest neighbors, which is included in dispersion law. If *N=1* we have harmonic dispersion law, so-called tight-binding approximation. It can be realized in SL with weakly-coupled well (wide barriers).

---

[*] e-mail: jul@ipm.sci-nnov.ru

Let us consider electric field of the type

$$E(t) = E_C + \sum_{\alpha=1}^{M} E_\alpha \cos(\omega_\alpha t - \delta_\alpha), \quad (2)$$

having dc component $E_c$ and *M* harmonics with frequencies $\omega_\alpha = n_\alpha \omega_1$ and beginning phases $\delta_\alpha$. We assume that the electric field is directed along the SL axis. In what follows, we drop the subscript "3" from longitudinal energy only if this does not lead to confusion.

Under influence of the electric field Eq.(2) an electron executes nonlinear oscillations having velocity given by:

$$V(k_0,t_0,t) = \sum_{\nu=1}^{N} V_m^\nu \sin\left[\theta_\nu + \nu\Omega_c t + \nu\sum_{\alpha=1}^{M} g_\alpha \sin(\omega_\alpha t - \delta_\alpha)\right], \quad (3)$$

$$\theta_\nu = \nu\left[k_0 d - \Omega_c t_0 - \sum_{\alpha=1}^{M} g_\alpha \sin(\omega_\alpha t_0 - \delta_\alpha)\right], \quad (4)$$

where $V_m^\nu = \Delta_\nu \nu d / 2\hbar$ $V_m = \Delta d/2\hbar$ is maximum longitudinal velocity in *ν-th* partial minizone, $k_0$ is the electron wave vector at initial time $t_0$ (hereinafter we suppose $t_0 = 0$, $\delta_1 = 0$), $\Omega_C = eE_C d / \hbar$ is the Bloch frequency, $g_\alpha = eE_\alpha d/\hbar\omega_\alpha$ are dimensionless amplitudes of fields harmonics. Every *ν-th* partial minizone excites a partial Bloch oscillation with the frequency $\nu\Omega_c$, which makes additive contributions to whole Bloch oscillation. If $\Omega_c = \lambda\omega_1$ the time averaged electron velocity is given by

$$\overline{V(k_0,t_0,t)} = \sum_{\nu=1}^{N} \overline{V_\nu},$$

$$\overline{V_\nu} = -\frac{i}{2} V_m^\nu \exp(-i\theta_\nu) \sum_{\mu_2...M=-\infty}^{\infty} J_{-\lambda\nu - \sum_{\alpha=2}^{M} n_\alpha \mu_\alpha}(\nu g_1) \prod_{\beta=2}^{M} J_{\mu_\beta}(\nu g_\beta) \cdot \exp\left[-i\sum_{\alpha=2}^{M} \mu_\alpha \delta_\alpha\right] + c.c.,$$

(5)

Thus the DL conditions will be such:

$$\sum_{\mu_2...M=-\infty}^{\infty} J_{-\sum_{\alpha=2}^{M} \mu_\alpha n_\alpha - \nu\lambda}(\nu g_1) \cdot \prod_{\alpha=2}^{M} J_{\mu_\alpha}(\nu g_\alpha) \cdot \cos\left(\sum_{\beta=2}^{M} \mu_\beta \delta_\beta\right) = 0,$$

$$\sum_{\mu_2...M=-\infty}^{\infty} J_{-\sum_{\alpha=2}^{M} \mu_\alpha n_\alpha - \nu\lambda}(\nu g_1) \cdot \prod_{\alpha=2}^{M} J_{\mu_\alpha}(\nu g_\alpha) \cdot \sin\left(\sum_{\beta=2}^{M} \mu_\beta \delta_\beta\right) = 0, \quad \nu = 1...N.$$

(6)

The second Eq.6 for odd $n_\alpha$ and $E_c=0$ is satisfied automatically. Such situation repeated for any $n_\alpha$ and $E_c=0$ if a phase shift is equal to $\delta_\alpha=0, \pm\pi$.

It is easy to see that the opportunity of DL occurrence depends on ratio between number of harmonics in dispersion law and in electric field. Namely, if the field contains even harmonics the number of partial minizones (*N*) in dispersion law can be less than the number of electric field harmonics (*N<M* if *N>1*). Only in this case DL occurs in field which form is determined by Eq.6.

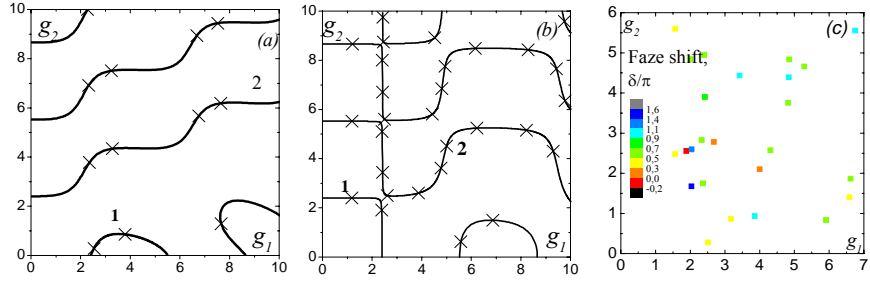

Fig1. The conditions for DL and SIT on the plane ($g_1,g_2$) for different approximations to the miniband structure of SL in biharmonic field. $\omega_2 = 3\omega_1$. N=1: a,b, solid lines; N=2: a,b - crosses, (a) - $\delta=\pi$, (b) - $\delta=\pi/2$; N=3: (c)

If electric field includes only odd harmonics requirements to dispersion law are less restrictive: N≤2M-1. In particular a biharmonic field can cause DL in SL with three or less partial minizones. Fig.1 represents DL conditions in biharmonic field ($n_2=3$) for different approximations to the band structures $\varepsilon(k)$.

If we add static field the situation changes appreciable. The static field shifts [1] the spectrum of each partial BO by the "partial" Bloch frequency $\nu\Omega_c$. Let us consider the static field with amplitude $\Omega_c =n_1/n_C\omega_1$ ($n_{1,C}$ are integer but have not common multiplier). Then the period of Bloch oscillations is $n_c$ times as many as the period of field ($T_b=2\pi n_c$). The zero harmonic is absent in the velocity spectrum if the SL dispersion law does not contain the harmonic with number $\nu=sn_c$ ($s=1,2$). Recently this semi-classical result was confirmed by quantum calculations [6]. In particular it means that static field with amplitude $\Omega_c =n_1/n_C\omega_1$ allows to observe DL in the SL with $(M+1)n_c-1$ or less partial minizones. It is almost in $n_c$ time lager then in periodic field. (We have to satisfy conditions (6) only for some partial minizones with $\nu=sn_c$.)

It is easy to obtain that SIT occurs at the same values of the fields amplitudes (within single $\tau$ - approximation ($\tau$=const) and $\omega_{1,2}\tau>>1$).

## 3. Induced and Self-Induced Transparency

The case $\delta_2=\pi/2$, $n_2=2l-1$, $l=1,2,..$ (Fig 1,b) is interesting. Within the DL conditions for the harmonic field $J_0(g_1^\lambda) = 0$, $\lambda$ is a number of Bessel function root) SL keeps transparency to the second field of the named type up to amplitudes $g_2<g_1$ if $g_1<<2n_2$ and $\lambda<l$. The exceptions are the points $J_0(g_2)\approx 0$. For $n_2=3$ transparency state arises at $g_1=2.405$ and for $n_2=5$ it arises at $g_1=2.405$ and $g_1=5.5$, for example. We can call this phenomenon "*induced transparency* in the multi-fold frequencies fields". The phase shift is important in this case in contrast to a case of fields with aliquant frequencies [7].

## 4. The Stability of Transparency States Equations

For investigation of the stability of transparency states we supplement the Boltzmann equation in the $\tau$ -approximation [4] with the equation of continuity of current. (Detailed scheme description is adduced in the Ref.4). Let us consider the SL with low electron concentration. In that case the intrinsic field differs from external one slightly. The collision is supposed to be rare ($\omega_1\tau>>1$).

There are three different scenario of SL behavior which is in transparency state under biharmonic field. Which process will dominate depends on ratio between values of energy losses on the different frequencies.

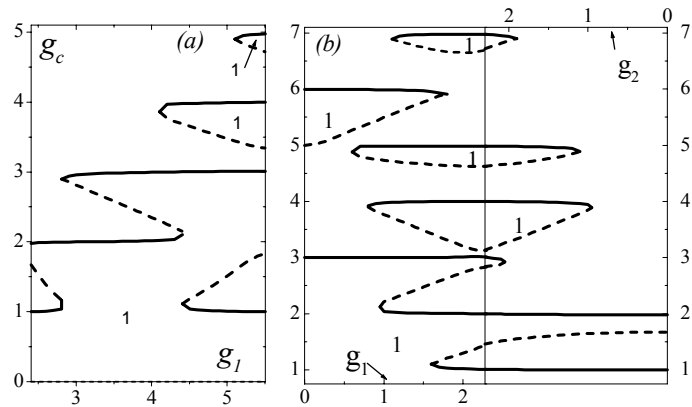

Fig.2. The regions of the absolute negative conductivity (1) of SL in a biharmonic field (Eq.2), $\omega_2 = 3\omega_1$ $E_c=0$). Value of $g_2$ was determined from Eq.6) (curves (1) on the Fig.1); $\omega_1\tau=10$. (a) $\delta_2=\pi$, (b) -$\delta_2=\pi/2$. The boundary curves of region (1) correspond to stable (solid line) and unstable (dash line) stage with $j_c =0$.}

(i) The destruction of the SIT state will be accompanied by spontaneous generation of dc field. Dc field value differs from the case of pure harmonic field (Fig.2).
(ii) The destruction of the SIT state will be accompanied by amplification of field and current harmonics with relatively large amplitudes.
(iii) The stabilization of SIT will be accompanied by generation of relatively week current and field harmonics.

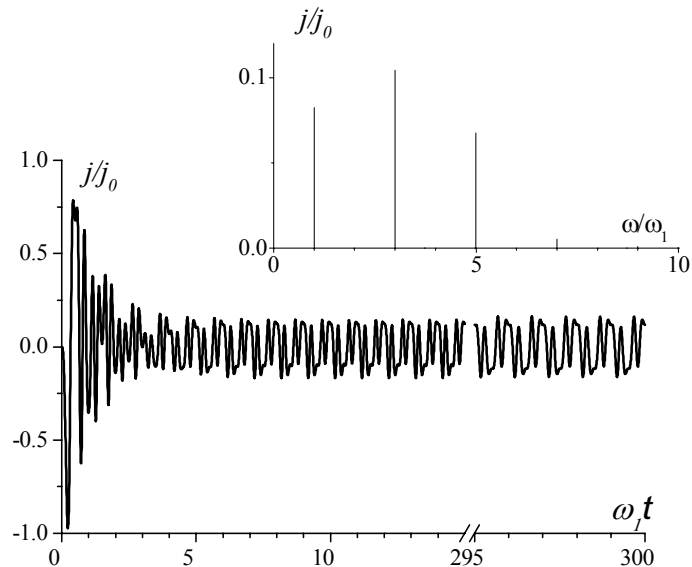

Fig.3. A stationary transparency state of a SL with a low electron concentration exposed to a biharmonic field ($n_0=3$, $\delta_2=\pi$). Time evolution and spectra of the current (in the insets). $\omega_1\tau=10$. $g_1=4.97$, $g_2=0.478$.

The last process will suppresses both dc field generation and amplification of components of biharmonic field (Fig.3). This stabilization can be stimulated by addition of corresponding weak harmonic into external field. There are periodic fields with complicated time dependance that form stable electrodynamic transparency states.

## 5. Conclusion

The conditions for dynamic localization and electromagnetic transparency in multifrequency electric field were founded,

The induced transparency were shown to be arise in biharmonic field with multifold frequencies($\omega_2 = n_2\omega_1$, $n_2$ is odd) under certain conditions.

In common case the transparency states are unstable. It is possible to stabilize them in multifrequency fields.

**Acknowledgments**

This study was supported by the Russian Foundation for Basic Research (projects no.05-02-17319).